%% file: main.tex
\documentclass{egpubl}
\usepackage{eg2024}

\ConferencePaper        
\CGFStandardLicense

\usepackage[T1]{fontenc}
\usepackage{dfadobe}  
\usepackage{booktabs} 
\usepackage[normalem]{ulem}
\usepackage{subfig}
\usepackage{wrapfig}
\usepackage[pdftex]{graphicx}
\usepackage{pgfplots}
\usepackage{amsmath,amssymb}
\usepackage{multirow}
\usepackage{float}

\newcommand{\REV}[1]{\textcolor{blue}{#1}}

\usepackage{xspace}
\newcommand{\ie}{\textsl{i.e.}~}
\newcommand{\eg}{\textsl{e.g.}~}

\biberVersion
\BibtexOrBiblatex
\usepackage[backend=biber,bibstyle=EG,citestyle=alphabetic,backref=true]{biblatex} 
\electronicVersion
\PrintedOrElectronic

\ifpdf \usepackage[pdftex]{graphicx} \pdfcompresslevel=9
\else \usepackage[dvips]{graphicx} \fi

\usepackage{egweblnk} 

\title[Cinematographic Camera Diffusion Model]
{Cinematographic Camera Diffusion Model}

\author[\centering H. Jiang, X. Wang, M. Christie, L. Liu \& B. Chen]
{
\parbox{\textwidth}{\centering Hongda Jiang$^{1,3}$\orcid{0000-0002-0296-4431}, Xi Wang$^{4}$\orcid{0000-0001-6586-1926}, Marc Christie$^{5}$, Libin Liu$^{2,3}$\orcid{0000-0003-2280-6817} and Baoquan Chen\thanks{Corresponding author
}$^{2,3}$} \\
{\parbox{\textwidth}{\centering $^1$School of Computer Science, Peking University\\
$^2$School of Intelligence Science and Technology, Peking University\\
$^3$National Key Laboratory of General AI\\
$^4$Team VISTA, LIX, École Polytechnique\\
$^5$University Rennes, Inria, CNRS, IRISA}}
}

\begin{document}

\teaser{
 \includegraphics[width=\linewidth]{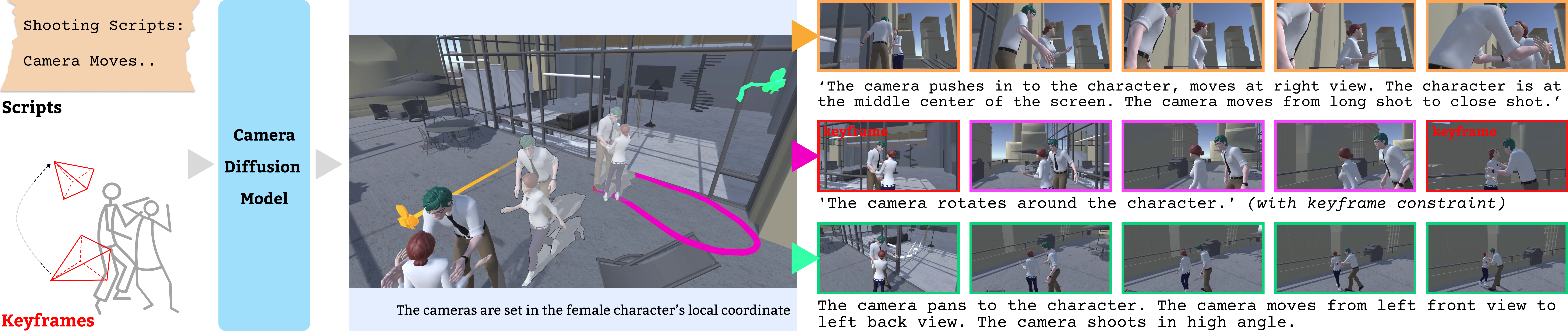}
 \centering
  \caption{We propose the design of a camera diffusion technique that generates diverse and realistic camera motions conditioned on cinematographic textual inputs and optional keyframe constraints (red camera icons on the left). The model provides users with a natural and creative way to build camera motions and blend the motions while enabling precise control using keyframes.}
\label{fig:teaser}
}

\maketitle
\begin{abstract}
Designing effective camera trajectories in virtual 3D environments is a challenging task even for experienced animators. Despite an elaborate film grammar, forged through years of experience, that enables the specification of camera motions through cinematographic properties (framing, shots sizes, angles, motions), there are endless possibilities in deciding how to place and move cameras with characters. Dealing with these possibilities is part of the complexity of the problem. While numerous techniques have been proposed in the literature (optimization-based solving, encoding of empirical rules, learning from real examples,...), the results either lack variety or ease of control.

In this paper, we propose a cinematographic camera diffusion model using a transformer-based architecture to handle temporality and exploit the stochasticity of diffusion models to generate diverse and qualitative trajectories conditioned by high-level textual descriptions. We extend the work by integrating keyframing constraints and the ability to blend naturally between motions using latent interpolation, in a way to augment the degree of control of the designers. We demonstrate the strengths of this text-to-camera motion approach through qualitative and quantitative experiments and gather feedback from professional artists. The code and data are available at \URL{https://github.com/jianghd1996/Camera-control}.

\ccsdesc[500]{Computing methodologies~Procedural animation}
\ccsdesc[500]{Computing methodologies~Artificial intelligence}

\printccsdesc   
\end{abstract}  

\section{Introduction}
\input{1_introduction}

\section{Related Work}
\input{2_related-work}

\section{Overview}
\input{3_overview}

\section{Camera Diffusion Model}
\input{4_methods.tex}

\section{Results and Experiments}
\label{sec:evaluation}
\input{5_results.tex}

\section{Limitations and Conclusion}

\input{7_conclusion}

\section*{Acknowledgement}
\input{9_ack}

\printbibliography

\appendix
\input{10_appendix}

\end{document}

%% file: 1_introduction.tex




\textit{"It’s only through writing scripts that you learn specifics about the structure of film and what cinema is.” -- Akira Kurosawa}

Films are not only about moving pictures, but are also created using language. They are produced through the dialogues spoken by the characters, the screenplays that depict the narrative, the shooting scripts that instruct the cinematographers, and, undeniably, the guidance given by directors. During the filmmaking process, filmmakers communicate their inspiration through verbal or written linguistic commands, instructing cinematographers to perform various artistic camera movements that help convey the story and express emotions within the image frames.

The language used to discuss cinematography is not confined to a prestige language that serves only experts. With the rapid development of realistic video games, cameras accessible on mobile phones, and even unmanned photography robots such as drones, the language of cinematography is becoming increasingly prevalent among the general public. 
However, converting such high-level descriptions to specific camera placements and motions in real or virtual worlds raises several challenges: i) the regular user may not be able to position the camera with expertise, given the large implicit knowledge in cinematography, \eg confirming the possible range of camera positions to well-established frame scales, angles, compositions, and intended emotional and cinematic messages. For instance, \textit{"pushing in to a close-up shot"} typically refers to implicit specific motions and camera velocities with ease-in and ease-out effects, spatial distances, and on-screen framing properties with respect to the character's facial area on screen; ii) film language also exhibits many ambiguities, as in every human spoken language and the same linguistic input could refer to a range of different plausible results. In the traditional filmmaking workflow, cinematographers and directors of photography are responsible for dealing with such ambiguities, interpreting operational camera motions from written screenplays and camera shooting scripts (the language command for the cinematographer).

On the computer science side, diffusion models have recently demonstrated significant advancements in the quality, manipulation, and versatility of AI-generated content (AIGC). Applications encompass a wide range of topics, including image and video generation~\cite{ramesh2022hierarchical}, human animation~\cite{tevet2022human}, and even 3D content generation~\cite{poole2023dreamfusion}. Many of these approaches rely on CLIP text embedding~\cite{radford2021learning} as guiding information to condition the output generation from human linguistic input. For instance, DALLE~\cite{ramesh2022hierarchical} enables users to describe an image in terms of content and style. Similar applications can be observed in the computer animation field such as the Human Motion Diffusion Model (MDM)~\cite{tevet2022human}, Speech Gesture Diffusion~\cite{ao2023gesturediffuclip}, etc.

The question we address in this work is whether we can leverage such diffusion models to build a script-to-camera-motion approach for virtual cinematography that would enable the generation of diverse results while also encoding implicit cinematographic characteristics. Unlike image and animation generation, the use of natural language for cinematography poses unique challenges and the following demands should be addressed:

\noindent
\textbf{Cinematic paradigm \emph{vs.} linguistic ambiguities:} The generated results shall align with established cinematographic principles (\ie the large implicit knowledge composing the cinematic paradigm~\cite{albera2012}) while being able to handle linguistic ambiguities. This necessitates the ability to understand the implicit cinematographic knowledge and be able to propose a \emph{many-to-many} generative system addressing \emph{many-to-one} and  \emph{one-to-many} challenges. {The \emph{many-to-one} challenge is rooted in the linguistic variations that arise from different users when giving commands to the camera system, \eg 'push in', 'get closer', and 'move towards target' can represent similar camera motions. Additionally, a camera shot can be articulated from various perspectives, encompassing aspects such as camera movement, shot composition, and screen properties. The \emph{one-to-many} challenge is more grounded on the ambiguous and partial nature of such a language. For example, a 'tracking shot' can be executed from different directions: from back view, side view, or front view, therefore opening rich possibilities for designers.}

\noindent
\textbf{User controllability \emph{vs.} style constraint:} In the context of camera motion generation, user controllability plays a critical role, particularly in filmmaking scenarios where precise camera movements are required to avoid obstacles and achieve specific shots, whereas the common text-based condition may not be sufficient to address such requirements. To tackle this challenge, it is necessary to address the common \textit{keyframing} and \textit{transitioning} problems to enable enhanced controllability without compromising the generative capacity of cinematographic style constraints. The seamless transition between sequences of generated results is also crucial to ensure smooth and artifact-free outputs.

In this context, the objective of our work is to generate \textbf{diverse and realistic} camera motions from textual descriptions and offer additional control to the users by specifying keyframe camera conditions and means to transition realistically between different motions. To this end, we propose to rely on a diffusion approach that we extend to handle camera motion specificities (automated transitions and keyframe control). Different from regressive models such as LSTM, generative models like GANs (generative adversarial networks) and diffusion models have demonstrated impressive capacity in learning distributions and synthesizing unseen and diverse results. GANs, however, hold shortages \eg: difficulty in training without carefully selected parameters, complicated when enabling conditioning~\cite{dhariwal2021diffusion}. Furthermore, results seem to demonstrate that, despite similar performance on FID (Frechet Inception distance, which measures similarities between distributions) or on precision, GANs capture less diversity and are more prone to mode dropping than state-of-art likelihood-based models. This is because while training the generator fails to sample from the full range of possibilities in the target distribution, causing it to produce samples that are too similar to each other~\cite{dhariwal2021diffusion,xiao2021tackling}.

Therefore, drawing inspiration from motion diffusion models~\cite{tevet2022human}, we present our method named Camera Diffusion Model (CDM), which learns from text-driven CLIP embeddings how to generate diverse and realistic motions. We benefit from the \emph{one-to-many} capacity thanks to the stochasticity and better mode coverage of diffusion models, and extend the \emph{many-to-one} capacity through textual augmentations. Furthermore, in order to integrate additional constraints in the specifications, we use keyframe constraints as input conditions, rather than exploiting traditional inpainting techniques. Finally, we exploit the interpolation capacity of the CLIP embedding to demonstrate how different generated sequences can be smoothly blended, while remaining qualitative.




The contributions of our work are the following:\\
\noindent\textbf{(i)}. To the best of our knowledge, we are the first to design a cinematographic camera diffusion system that incorporates linguistic support and provides keyframing constraints together with realistic transition schemes between camera sequences. This enables users to generate diverse camera motions for given animations, even when facing linguistic ambiguities. It also ensures that our generative model is a practical tool for artists and non-experts alike, providing both precise control and the ability to generate cinematically pleasing results.

\noindent
\textbf{(ii)}. We show that these additional controls though keyframing and transitioning schemes address the inherent conflict between controllability and styled generation. We empirically illustrate that prior inpainting techniques using masking~\cite{tevet2022human} fail to handle this conflict and demonstrate the superiority of integrating keyframes as extra condition information in the diffusion process.

\noindent
\textbf{(iii)}. We finally conduct extensive quantitative and qualitative experiments, ablation studies, and showcases on various aspects of our method, including the impact of linguistic certainty, the trade-off between keyframing and style, zero-shot multimodality inference, etc.

The source code and dataset for our method can be found at \URL{https://github.com/jianghd1996/Camera-control}. We hope our work can provide valuable resources for further exploration and experimentation in this field.

%% file: 2_related-work.tex
\textbf{Cinematography \& camera control}.  
The field of computational cinematography covers a wide spectrum of problems, from narrative aspects~\cite{de2009virtual} to mise-en-scene~\cite{xu2002constraint}, lighting~\cite{shacked2001automatic}, camera control~\cite{christie2008camera,gleicher1992through,li2008real} and even cutting and editing problems~\cite{leake2017computational}. Many researchers also worked on more constrained problems such as drone cinematography~\cite{galvane2018directing,nageli2017real}. In recent years, the field has embraced Deep Neural Networks (DNNs) due to their remarkable fitting capacity and ability to generalize from complex datasets. Applications are found in cinematic feature extraction~\cite{courant2021high}, film and shot classification and segmentation~\cite{rao2020local}, cutting and montage~\cite{wang2019write,Pardo_2021_ICCV}, film scene reconstruction~\cite{angjoo_TV_show}, etc.

Furthermore, several deep learning-based systems have been developed to address the camera control problem in various contexts: such as drone cinematography systems~\cite{huang_drone, bonatti2020autonomous}, which propose practical solutions for drone shooting tasks but are limited in learning only certain types of camera movements and lack fine-grained control. In the domain of virtual cinematography, \ie applying cinematography in computer graphics environments, Jiang et al.~\cite{hongda_example} combined the Toric~\cite{toric} coordinate system with a Mixture-of-Experts to generate styled camera motions based on different video reference inputs. A follow-up work ~\cite{hongda_keyframe} introduced keyframing for finer control of camera motions using an LSTM-based backbone. \REV{These works design camera generators with examples and keyframe control, which are not effective for textual conditions.} Additionally, in alternative implicit learning-based environments such as NeRF~\cite{mildenhall2021nerf}, \cite{wang2023jaws} proposed a cinematic transfer task aimed at replicating similar camera motions from a reference video within a NeRF scene.

\noindent
\textbf{Generative models}. One approach to the computational cinematography task involves framing it as a generative problem with a predefined data distribution. However, the technical approaches to solving the generative problem are diverse in deep learning research: from Auto-Encoder, Variational Auto-Encoder~\cite{kingma2013vae} to the popular GAN~\cite{goodfellow2020gan} models. Recently, a contender has emerged in the longstanding competition in generative models: the diffusion model, initially proposed in~\cite{sohl2015deep} and becoming prevalent in the image generation domain due to the adaptation of DDPM~\cite{ho2020denoising}. The model represents a notable advancement in this field. The main idea consists of reversely generating content from Gaussian noise latent representation step by step and has demonstrated remarkable capabilities in synthesizing realistic image content. Subsequently, enhancements have been made to improve user controllability and generation quality through the utilization of guidance such as classifier guidance~\cite{classifier_guidance} and classifier-free guidance~\cite{ho2022classifier}. 

\noindent
\textbf{Text-driven generative models}. Many recent works have chosen to leverage textual inputs or conditions to drive generative models. These approaches have the potential to democratize technologies and benefit the rapid development of Large-Language Models (LLMs) for facilitating tedious and repetitive tasks. One of the foundational works in this area is the CLIP model~\cite{radford2021learning} that exploits contrastive learning to create a joint image-text embedding space. CLIP has demonstrated excellent performance on zero-shot image and text tasks. This seminal work has opened up numerous possibilities for incorporating text commands into the generative process, allowing users to linguistically describe the desired content to be generated. The very combination between the controllable CLIP model and the generative diffusion model empowered many applications in computer vision and computer graphics domains, enabling text manipulation on generated content of images~\cite{ramesh2022hierarchical, rombach2022high}, videos~\cite{ho2022video}, human motion~\cite{tevet2022human} as well as human gesture animation~\cite{ao2023gesturediffuclip}.

In our work, we build on the recent contributions related to human motion generation \cite{tevet2022human}, and we design dedicated means to handle the specific one-to-many and many-to-one requirements of text-driven cinematographic generation. We also propose means to handle keyframe conditioning for improved control and qualitative motion interpolation, which contrast with existing work.

%% file: 3_overview.tex
 The overall architecture of our system is described in Fig.~\ref{fig:main_fig}. The system takes as input a partial or complete textual description that follows well-established framing, angle, and camera motion properties, together with optional user-defined camera keyframes on the extremities of the trajectory. We use the CLIP embedding model~\cite{radford2021learning} to encode the textual descriptions that we concatenate with keyframe information. A latent noise $X_t$ is also input to the camera diffusion model (CDM) to generate an output noise $X_{t-1}$. The system outputs a camera motion sequence satisfying both textual specification and keyframe constraints by denoising the output $X_t$ through successive steps. The system is also designed to ensure a smooth transition from one sequence to another with different strategies for long sequence generation.

%% file: 4_methods.tex

\label{label:Method}

\begin{figure*}[ht]
    \centering
    \includegraphics[clip,width=1.0\textwidth]{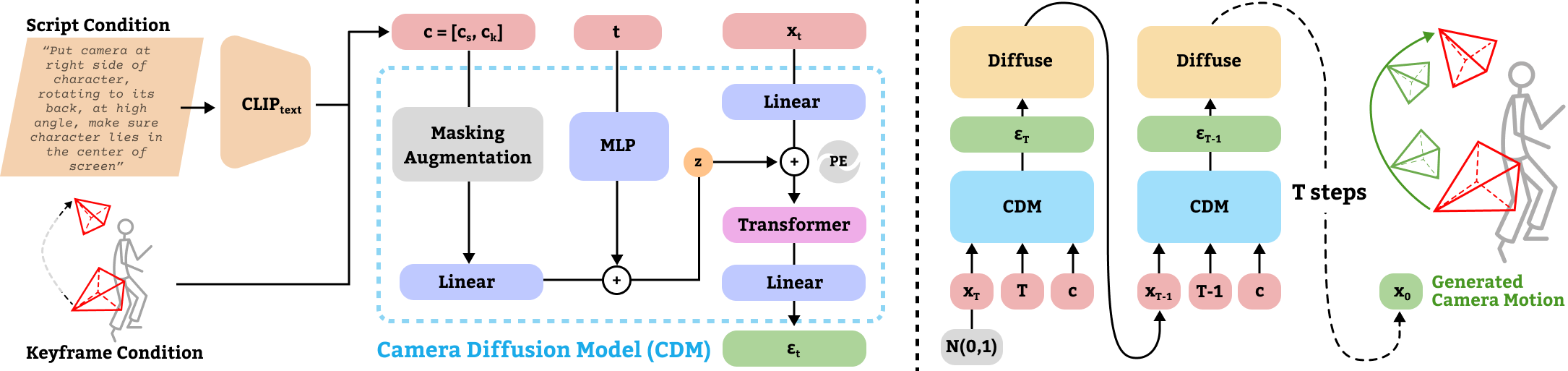}{\centering}
    \caption{The pipeline of the proposed system. Our proposed Camera Diffusion Model (CDM) takes script conditions, keyframes (optional), and a latent noise as input, and outputs camera motion sequence by denoising via the diffusion model $T$ steps. At each step $t$, CDM predicts the noise $\epsilon_t$ from $X_t$ and diffuses it to $X_{t-1}$.}
    \label{fig:main_fig}
\end{figure*}

\subsection{Data format} The proposed model aims to synthesize a sequence of camera poses $x_0^{1:N}$ (we replace the notation of $x_0^{1:N}$ by $X_0$ for simplicity) over $N$ frames with a given condition $c$. The camera pose at a given frame number $n$ is represented with character-centric local coordinates:
\begin{equation}
    X_0=\{x^n, y^n, z^n, p_x^n, p_y^n\}\in \mathbb{R}^5, n\in [1..N],
\end{equation}
where $\{x^n,y^n,z^n\}$ stands for the relative position of the camera in local character coordinates to the head center, and $\{p_x^n,p_y^n\}$ represents the normalized screen position of the 'Look-At' target (default head) at the frame $n$. We do not model the roll angle of the camera in this representation.



\subsection{Network design} 
The main idea behind the diffusion model is to gradually add increasing noise to the original data sequence $X_0$ and learn a denoising process in reverse steps, transforming the data from a noise latent to a valuable sequence. This stochastic process serves two purposes: i) effective fitting of the distribution of the original data; ii) handling the inherently ambiguous many-to-many mapping problem. By gradually removing noise and revealing signal, the diffusion model can generate multiple plausible outputs while not compromising on variability and diversity. This forward process can be described as:
\begin{equation}
    q(X_t|X_{t-1})=\mathcal{N}(\sqrt{\alpha_t}X_{t-1},(1-\alpha_t)I), t \in [1..T],
\end{equation}
where $\alpha_t \in (0,1)$ is a decreasing variable when the timestep $t$ gets larger given a certain noise scheduler. 
When $\alpha_t$ is small enough, the distribution of $X_t$ can be approximately viewed as a standard Gaussian distribution. Then, the text-based camera motion generation task can be cast as a denoising process from a randomly sampled Gaussian noise with a condition $c$, $p(X_0|c)=f(X_T|c)$ (see top of Fig.~\ref{fig:main_fig}). Similar to~\cite{ho2020denoising,tevet2022human}, using the notation $\bar{\alpha}_t:=\prod_{s=1}^t\alpha_s$, we adopted a simple training objective to learn the reversed noise with a random sampled noise $\epsilon \in \mathcal{N}(0,1)$ and estimated noise $\epsilon_{\theta}$ from network: 
\begin{equation}
    \mathcal{L}(\theta)=E_{t,X_0,\epsilon} [ || \epsilon-\epsilon_{\theta}(\sqrt{\bar{\alpha}_t}X_0+\sqrt{1-\bar{\alpha}_t}\epsilon, t)||^2 ].    
\end{equation}


There are two primary challenges when applying the diffusion model directly to camera cinematography tasks: (i) while the text-based condition allows for incorporating style in the generated camera motion, the level of controllability is insufficient for precise camera control scenarios often required by designers. In such cases, users typically require the possibility to create specific camera positions as constraints, similar to keyframing techniques, to control the overall sequence better; (ii) as the diffusion model generates results in a segment-to-segment fashion, ensuring continuous and smooth transitions between different segments is crucial for producing visually appealing camera motions.


\subsection{Keyframes as constraints in the diffusion process} 
Regarding keyframing strategies, many previous works have relied on inpainting~\cite{lugmayr2022repaint} or in-betweening schemes to impose additional constraints on the generated results. For example, MDM~\cite{tevet2022human} sets partial sequence information, such as the first 25\% and last 25\% of the content, as fixed during the denoising generation process. We implemented the inpainting scheme inspired by~\cite{lugmayr2022repaint},
\begin{equation}
    \begin{aligned}
    X_{t-1}^{\mathrm{keyframe}} &\sim  \mathcal{N}(\sqrt{\bar{\alpha}_{t}}X_{0},(1-\bar{\alpha}_{t})I),  \\
    X_{t-1} &= m\odot X_{t-1}^{\mathrm{keyframe}}+(1-m)\odot X_{t-1}^{\mathrm{predict}},
    \end{aligned}
    \label{eq:inpainting}
\end{equation}
where $X_{t-1}^{\mathrm{predict}}$ is predicted from last diffusion step and $X_{t-1}^{\mathrm{keyframe}}$ is two side keyframe conditions. These variables are combined to the new sample $X_{t-1}$ using the mask $m$. However, in this paper, we argue that this masking strategy is not efficient and fails to address the inherent conflicts between the style label and keyframing constraints. In the context of inpainting techniques, keyframe guidance may be regarded as noise in the denoising process due to its inherent sparsity. Instead, we incorporate the keyframing constraint $c_k$ directly as part of the condition.
We leverage the interpolation capacity of CLIP embedding and the keyframe condition design mentioned earlier to achieve continuous and styled transition segments between the generated results corresponding to different scripts. We show in the results that this joint approach enables us to generate visually coherent and cinematographically pleasing transitions in the camera motion sequences (see ablation studies in Section~\ref{subsec:keyframe_generation} for further discussion).

Therefore, to enhance the diffusion model with condition information $c$, we have employed the classifier-free guidance training protocol~\cite{ho2022classifier}. We train both an unconditional denoising diffusion model and a conditional model using a single neural network $\epsilon_{\theta}(X_t, t, c)$. During training, the condition is set to $\emptyset$ (unconditioned label) with a certain probability ($p=0.1$). During inference, the estimated noise is formulated as:
\begin{equation}
   \widetilde{\epsilon}_{t}=\epsilon_{\theta}(X_t, t, \emptyset) + \omega \cdot ( \epsilon_{\theta}(X_t, t, c)-\epsilon_{\theta}(X_t, t, \emptyset)).
\end{equation}
Our conditions include both cinematic shooting script condition $c_s$ and keyframe constraints $c_k$, aiming to cover both style guidance and precise specification of trajectory: $c=[c_s,c_k]$. For the shooting script, we rely on CLIP~\cite{radford2021learning} text embedding to map the text into a latent embedding $c_s\in \mathbb{R}^{512}$ to capture the semantic information and incorporate it into the diffusion model (Fig.~\ref{fig:main_fig} left). Regarding the keyframe condition, we encode the starting and ending keyframes into a representation $c_k \in \mathbb{R}^{2 \times 5}$, to provide more precise control to users. We train the network both with ($c=[c_s,c_k]$) and without keyframes ($c=[c_s,\emptyset]$) with the same probability.

\subsection{Dataset and training details}

\subsubsection{Synthetic dataset}

We generate a synthetic dataset consisting of camera trajectories and paired textual descriptions. The dataset encompasses variations in shot angle, shot scale, camera movement, camera velocity, and screen properties. The definitions of each of these terms are provided below. Please see the appendix~\ref{appendix:data_gen} for more details.

\noindent
\textbf{Shot angle.} The shot angle refers to the relative vertical angle between the camera and the target, which results in different visual perceptions. For example, a low angle is typically used to create a sense of grandeur or threat towards the subjects in the shot, while a high angle can convey vulnerability or insignificance. 

\noindent
\textbf{Shot scale.} The shot scale refers to the size of the target within the camera framing. We adopt a default field of view 45, and determine the shot scale based on the appearance of the characters or subjects within the screen~\cite{arijon1991grammar}. For example, in a medium shot, the bottom of the screen aligns with the characters' knees. We generate a range of camera shots from \textit{extreme close-up} to \textit{extreme long shot}.

\noindent
\textbf{View directions.} The view direction describes the relative position between the camera and the target. We enumerate eight simple directions to represent different perspectives: \textit{front, back, left, right, left front, left back, right front, right back}.

\noindent
\textbf{Screen properties.} Instead of using a camera-centric orientation, we leverage the screen position of the target to control the camera's orientation. Following the common screen composition rules \cite{arijon1991grammar}, we define the screen position of the target as a relative vertical and horizontal position, resulting in combinations of \textit{top-middle-bottom} and \textit{left-center-right} on the screen.

\noindent
\textbf{Shot movement.} We describe several basic camera movements with character-centric coordinates: \textit{Static}: The camera remains locally static in both position and framing; \textit{Push in/Pull out}: The camera either decreases or increases the distance to the character, while keeping other properties constant; \textit{Pan}: The camera position remains constant while it rotates on its vertical axis to track a target or switch between targets. In character-centric coordinates, we transform the globally static camera into a locally moving camera; \textit{Boom}: A boom motion typically involves an upward or downward translational movement of the camera. This camera movement is commonly used at the beginning or end of a narrative; \textit{Orbit}: In this motion, the camera swivels horizontally around a target. Here, we randomly select the starting and ending view directions, as well as a moving direction, to generate a circular movement with a controlled angular velocity. This motion is widely used in modern films for various reasons: engaging dynamic motion, navigating around the environment, or highlighting the shot target.

We collect 25,000 sequences of data, amounting to a total of 4,500,000 frames, together with 206,570 single sentences of textual descriptions. By default, the length of all the synthetic data is limited to 300 frames. However, to introduce variations in camera velocities, we further augment the camera movement by scaling the time factor in the sequences. For instance, a \textit{fast} push-in movement may last approximately 3 seconds, while a \textit{slow} push-in movement may last around 8 seconds. We demonstrate how to control the camera velocity during inference in Section~\ref{label:vel_control}.

We create our dataset by combining variations across all properties. Different camera movements exhibit specific variations in each property, e.g. the \textit{push in/pull out} shot has different shot scales at the beginning and ending frames. We enumerate all possible combinations and randomize the parameters within the specific range, resulting in a collection of primitive camera sequences.





\subsubsection{Prompt augmentation}

We utilize the CLIP~\cite{radford2021learning} model to generate embeddings as conditions from text prompts. We propose variant template prompts to describe different camera properties, where each property is generated separately. For example, '\textit{The camera shoots the character in front view.}', '\textit{The character is at the middle center of the screen.}' '\textit{The camera pushes in to the character.}' Also, we augment the prompt with synonyms or similar expressions, such as describing the orbit movement as '\textit{the camera \textbf{rotates} around the character.}' and a static tracking in the side or back view as \textit{'the camera \textbf{tracks} the character'}. The full description of a camera sequence is the combination of all the prompts.

During the inference stage, users may provide few or partial descriptions of the desired camera motion. To handle such scenarios, we randomly select a portion of the given prompts during training as the condition to generate the camera motion sequence. This approach transforms the task into a challenging many-to-many generation problem, yet the diffusion model can generate multiple diverse results that all satisfy the partial given descriptions. \eg many different orbiting frequencies to one description of \textit{rotating}.

This operation provides users with a range of linguistic controls to accurately specify their desired camera motion gradually. By providing more detailed prompts step by step, the specification of the generated results can be more refined, resulting in less variance across different generations with the sampling noise. The trade-off between variation and controllability will be further discussed in the forthcoming Section~\ref{subsec:evaluation}.

\subsubsection{Training details}


We follow the DDPM~\cite{ho2020denoising} to train the diffusion model. 
During training, we utilized a single Titan Xp GPU with diffusion noising steps T=1000 and a linear noise schedule. The batch size is 256 and we set a learning rate of $1.0e{-4}$ for the Adam optimizer which follows linearly a reduction scheduler. We train our text condition with a respective maximum sequence length of 60 frames and a maximum sequence of 300 frames to satisfy different cases such as creating short transitional sequences or long sequences. With the fixed length requirement in the transformer, we employ padding for terminal frames when the sequence falls short of the desired target length. The training takes about 60 and 90 hours respectively to converge. During the inference phase, synthesizing a sequence of 300 frames requires around 6 seconds. Multiple sequences can be generated in parallel with conditions respectively. {We evaluate the model with 60 frames in Section~\ref{sec:evaluation} and employ both models (60 and 300 frames) to generate our results}.


\subsection{Inference} At inference time, we sample $p(X_0|c)$ from a standard Gaussian noise and then predict and remove noise from step $N$ to $0$ (Fig.~\ref{fig:main_fig} bottom right). The final conditioned results $f_c(X_t,t,c)$ can be computed by incorporating both conditioned generation $f(X_t,t,c)$ and unconditioned generation $f(X_t, t, \emptyset)$ shown as Eq.~\ref{eq:classifier_gen}. The balance of diversity and fidelity can be controlled by different magnitudes $\omega$, which is further studied in Sect.~\ref{subsection:omega} (see~\cite{ho2022classifier} for more details):
\begin{equation}
    f_c(X_t, t, c) = f(X_t, t, \emptyset) + \omega \cdot (f(X_t,t,c)-f(X_t,t,\emptyset)).
    \label{eq:classifier_gen}
\end{equation}

%% file: 5_results.tex
\subsection{Script-to-camera}
\label{subsec:script-to-camera}
In the \textit{script-to-camera} task, our method generates camera motions only based on given scripts as conditions, where the keyframing condition is masked. To demonstrate the capability of generating qualitative content, we assess the performance on various metrics:

\textit{R Precision classification (R Prec):} This metric is used to confirm the model's ability to generate content in the desired style. We trained the classifier with a multi-layer transformer encoder as a feature extractor and additional fully connected layers for classification. The architecture is described in appendix~\ref{appendix:network}.
The training and testing datasets are randomly separated in a 9:1 ratio with six different \textit{camera movements} as labels. In testing, we generate sequences conditioned on the text features and the R precision evaluates the classification accuracy of generated sequences.

\textit{Fréchet Inception Distance (FID):} FID is a common metric in generative models for revealing the similarity between the distributions of real and generated content. In computer vision tasks, it is typically computed by a pre-trained Inception-v3 model~\cite{szegedy2016rethinking}. A similar idea of projecting animation to features and comparing the FID in the feature space can also be found in many human motion models~\cite{tevet2022human}. Therefore, we leverage the middle-level features of the trained classifier to compute the FID metric with 12,000 sampled sequences by randomly selecting ground truth text features.

\textit{Diversity (Div) \& MultiModality (MM):} Similar to~\cite{tevet2022human}, we compute the Diversity metric, to measure the variability in the resulting motion distribution. With generative models, the objective is to closely approximate the distribution of real data to confirm the believability of the generated content. Additionally, we calculate Multimodality as the average variance from the same text prompt condition to demonstrate the model's capacity to generate one-to-many results from inherently ambiguous linguistic inputs. In practice, we sample tens of sentences and sample 100 trajectories for each prompt. The MM value is the mean value of diversity with each prompt.

For the sake of comparison, we run our system side-by-side against a SoTA keyframe-driven style-to-camera system~\cite{hongda_keyframe}. Such a system encodes an example sequence to a latent style code with a gating network and generates a camera sequence through generative experts that account for both the encoded style code and keyframe constraints. To perform our comparison (see Tab~\ref{tab:script-to-camera}), we first took ground truth camera sequences as inputs to the SoTA system (LSTM+Gating~\cite{hongda_keyframe}), to serve as a reference. Then for the other models, we used the textual descriptions linked to the ground truth sequences. And we compared the techniques by first, proposing to eliminate the gating network in LTSM+Gating and substituting it with the style code from the CLIP text embeddings (LSTM+CLIP), before retraining their network. With only the style input in script-to-camera evaluation, we exclude the keyframe constraints in both the LSTM~\cite{hongda_keyframe} and the diffusion model to ensure a fair comparison (see the comparison with keyframes in the next section). We then run variants with CLIP and T5~\cite{raffel2020exploring} embeddings.

The results of both methods are presented in Tab.~\ref{tab:script-to-camera}, with metrics computed on samples from the real data (namely test ground truth data) distribution. We mask the keyframe condition when keyframe constraints are absent (where $c=[c_s,\emptyset]$), and trained each model using two different approaches: \textit{1-to-1}, where the condition of the text prompt and data are injectively mapped, and \textit{1-to-N}, where we introduced the skill of prompt augmentation by selecting a portion of the text prompts during training to generate inherently ambiguous data pairs.


\begin{table}[htp!]
    \centering
    \caption{Evaluation of methods on the script-to-camera task. \textbf{Bold} indicates best result, $\rightarrow$ indicates that closer to real is better.}
    \resizebox{0.48\textwidth}{!}{
    \begin{tabular}{l|c|c|c|c}
        \hline
            Method & FID$~\downarrow$ & R Prec(\%)$~\uparrow$ & Div $\rightarrow$ & MM $\uparrow$\\
        \hline
         Real & 0.003 & 99.26 & 63.14 & - \\
        \hline
         LSTM+Gating ~\cite{hongda_keyframe} & 26.50 & 98.24 & 62.23 & - \\
         \hline
         LSTM+CLIP (\textit{1-to-1}) & 137.32 & 63.56 & 60.63 & - \\
         LSTM+CLIP (\textit{1-to-N}) & 97.96 & 88.35 & 61.67 & - \\
         Diffusion+T5 (\textit{1-to-N}) & 52.43 & 91.40 & 61.85 & 43.29 \\
         Diffusion+CLIP (\textit{1-to-1}) & 92.83 & 68.22 & 61.52 & 40.83 \\
         Diffusion+CLIP (\textit{1-to-N}) (Ours)  & \textbf{48.25} & \textbf{97.78} & \textbf{61.93} & \textbf{47.75} \\
        \hline
    \end{tabular}
    }
    \label{tab:script-to-camera}
\end{table}

As shown in Tab.~\ref{tab:script-to-camera}, the low FID and high R Prec values on real data confirm the effectiveness and fairness of the proposed classifier-based metrics. We observe that our proposed method (in both training schemes) outperforms the LSTM+CLIP methods on almost all metrics. The prompt augmentation enhances the network's ability to recognize textual input with increased robustness. The reference LSTM+Gating method~\cite{hongda_keyframe} attains optimal performance but exploits the full camera sequence as input (rather than text embedding) thereby being more effective in discerning style information. Our method significantly outperforms the LSTM+CLIP model, thereby highlighting the capacity to handle the inherent many-to-many challenges in textual input.


Furthermore, the metrics of Diversity and MultiModality highlight the benefits of prompt augmentation in improving the variability, believability, and overall performance of the generated camera motion. Prompt augmentation enables the generation of one-to-many results, which increases the diversity and multimodality of the generated camera motions. However, it should be noted that LSTM-based methods do not inherently support one-to-many generation, and as a result, the MM metric in the comparison table remains empty for these methods.

We also investigate the impact of different pre-trained textual encoders on the performance of the diffusion model. We conduct a comparative analysis between CLIP and a frequently employed T5 encoder~\cite{raffel2020exploring}. In contrast to the CLIP model, the pre-trained T5 module produces a sequence of features from textual input. Subsequently, we keep the T5 module in a frozen state and train an additional LSTM to aggregate the T5 output tokens, yielding the final text embedding. We choose the CLIP model as a textual encoder due to its superior capability to embed cinematic prompts. Nevertheless, the results with T5 embedding also outperform the LSTM-based method.

\begin{figure}[ht]
    \centering
    \includegraphics[clip,width=0.49\textwidth]{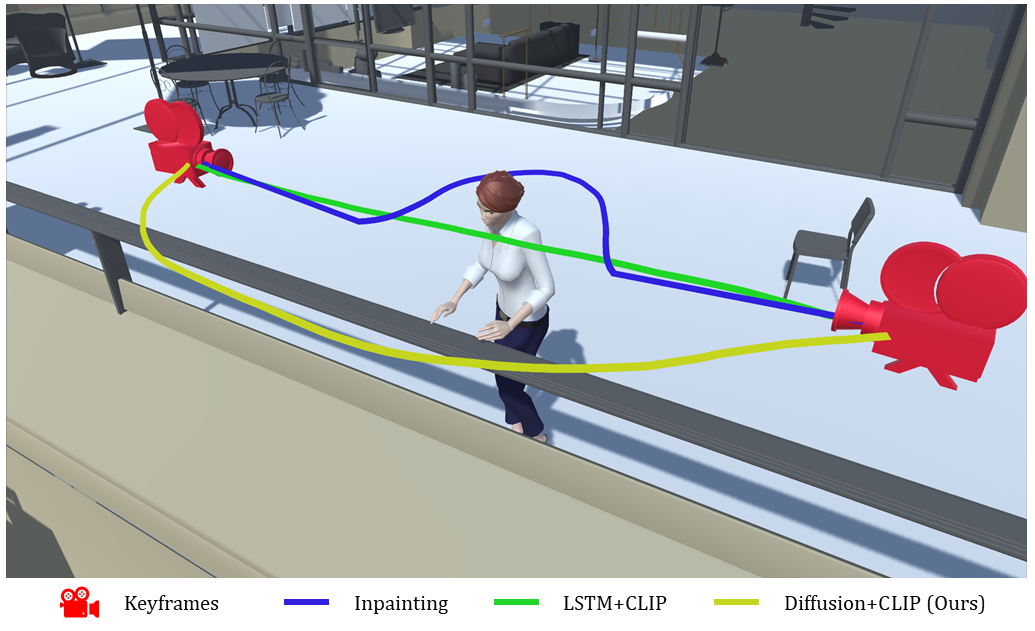}{\centering}
    \caption{{Comparison of generated trajectories with different methods (inpainting \emph{vs.} LSTM+CLIP \emph{vs.} Ours). Results are conditioned on the text prompt \textit{'The camera rotates around the character'}, and with the two keyframe constraints displayed in red.}}
    \vspace{-10px}
    \label{fig:keyframe_interp}
\end{figure}


\subsection{Keyframe-conditioned generation}
\label{subsec:keyframe_generation}

In this section, we present the experiments conducted when adding keyframing constraints by implementing and comparing different techniques: LSTM+Gating model~\cite{hongda_keyframe}, LSTM+CLIP model, keyframe conditioned diffusion generation (\textit{condition}) and inpainting based diffusion generation (\textit{inpainting}). As we mentioned before, inpainting techniques were proposed in \cite{tevet2022human,lugmayr2022repaint}. They consist in merely fixing some parts while applying the denoising process. We implement the inpainting technique in diffusion model with Eq.~\ref{eq:inpainting} using the text only condition $c=[c_s,\emptyset]$. The proposed \textit{condition} method incorporates both text and keyframes $c=[c_s,c_k]$ as conditions in denoising process.

{We evaluate different methods with text prompts from test dataset with keyframes close to the ground truth reference (near-style) and keyframes far to the ground truth (far-to-style) and display results in Tab.~\ref{tab:keyframe}}. It is important to note that far-to-style keyframes may intrinsically conflict with the desired style. \eg a keyframe turned way from the character may conflict with a push-in motion.


In addition to the FID and Precision metrics, we also introduce the keyframe distance (KF) metric to represent the spatial distance of the generated content in meeting the desired keyframes. We use the average L2 distance on \emph{the nearest 3 frames} of the starting and ending keyframes to the required keyframes to detect the degree of satisfaction of the keyframe constraint, as well as the continuity around keyframes (to avoid sudden jumps while still satisfying the first keyframe).

In Tab.~\ref{tab:keyframe}, we demonstrate that the \textit{condition} method outperforms the \textit{inpainting} method in generating believable content while satisfying the keyframing requirements and desired style. It is worth noting that 
{the utilization of the inpainting technique, while incorporating keyframe conditions during the denoising process, does not guarantee the generation of continuous trajectories. Employing the style-only classifier free guidance yields higher R precision, particularly in scenarios where keyframes may potentially conflict with the desired style.}



{Our method achieves comparable keyframe error with LSTM methods which incorporate an additional keyframe loss term during training. In far-to-style keyframe test, given the potential conflicts that may arise between the keyframes and the desired style, it is observed that the required style is not well maintained. We further investigate the style variation with the same groundtruth keyframe constraints. As shown in Tab.~\ref{tab:sameK_diffT}, we identify hundreds of pairs of camera trajectories sharing identical keyframes yet exhibiting distinct camera movements. It is evident that the learned LSTM module generates biased sequences, whereas our proposed method consistently provides a correct interpolation scheme.}

{To illustrate this, we provide an example in Fig.~\ref{fig:keyframe_interp} to elucidate the manner in which three distinct keyframe interpolation strategies generate camera sequences conditioned on two keyframes and an input text prompt. Notably, the LSTM+CLIP approach exclusively acquires the capacity to perform a linear interpolation during its training phase, while the inpainting method fails to align with the keyframes. In contrast, our method excels in generating accurate styles while adhering to the keyframe constraints.}

\begin{table}[h!]
    \centering
    \caption{Evaluation of near-style and far-to-style keyframes conditioned generation. In near-style keyframing (KF), the model is conditioned with keyframes close to the original data when generating the results. In far-to-style KF, keyframes are arbitrarily sampled, and the method is asked to generate camera motion with the desired style prompt.}
    \resizebox{0.48\textwidth}{!}{
    \begin{tabular}{l|c|c|c|c|c|c}
        \hline
            & \multicolumn{3}{|c|}{Near-style KF} & \multicolumn{3}{|c}{Far-to-style KF}\\
        \hline
            Method & FID$\downarrow$ & KF dist $\downarrow$ & R Prec $\uparrow$ & FID$\downarrow$ & KF dist $\downarrow$ & R Prec $\uparrow$\\
        \hline
         LSTM+Gating~\cite{hongda_keyframe} & 3.87 & 0.042 & 96.10 & 143.75 & 0.046 & 60.03 \\
        \hline
         LSTM+CLIP (\textit{1-to-1}) & 8.04 & 0.065 & 93.08 & 165.94 & 0.067 & 56.22\\
         LSTM+CLIP (\textit{1-to-N}) & \textbf{4.64} & 0.047 & 95.46 & 138.54 & 0.042 & 58.11 \\
        \hline
         Ours (\textit{inpainting}) & 43.14 & 1.794 & 79.83 & 114.53 & 1.95 & \textbf{63.30} \\
         Ours (\textit{condition}) & 6.67 & \textbf{0.041} & \textbf{96.62} & \textbf{111.03} & \textbf{0.041} & 58.05 \\
        \hline
    \end{tabular}}
    \label{tab:keyframe}
\end{table}

\begin{table}[h!]
    \centering
    \caption{{Evaluation of camera trajectories with the same keyframes but different camera movements.}}{
    \begin{tabular}{l|c|c}
        \hline
            Method & KF dist $\downarrow$ & R Prec $\uparrow$ \\
        \hline
         LSTM+CLIP (\textit{1-to-N}) & 0.048 & 50.0 \\
         Ours (\textit{condition}) & 0.039 & 92.8\\
        \hline
    \end{tabular}}
    \label{tab:sameK_diffT}
\end{table}

\begin{figure}[ht]
    \centering
    \includegraphics[clip,width=0.4\textwidth]{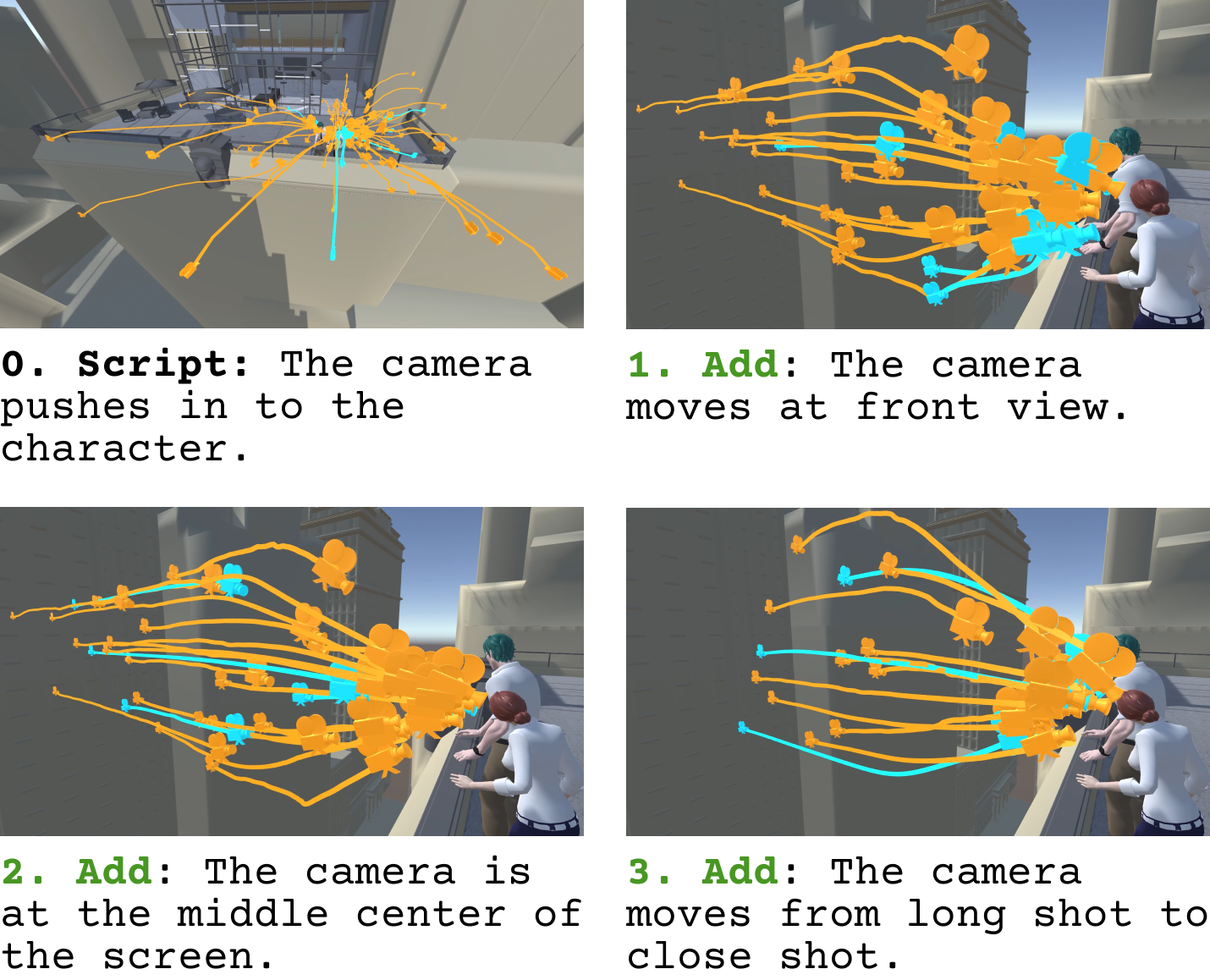}{\centering}
    \caption{By iteratively appending the descriptive information to the script, the generated motion progressively converges to the overall input condition. Highlighted cyan camera views are demonstrated in Fig.~\ref{fig:all_image_page1}.}
    \vspace{-10px}
    \label{fig:prompt_level}
\end{figure}


\begin{figure}[ht]
    \centering
    \includegraphics[clip,width=0.47\textwidth]{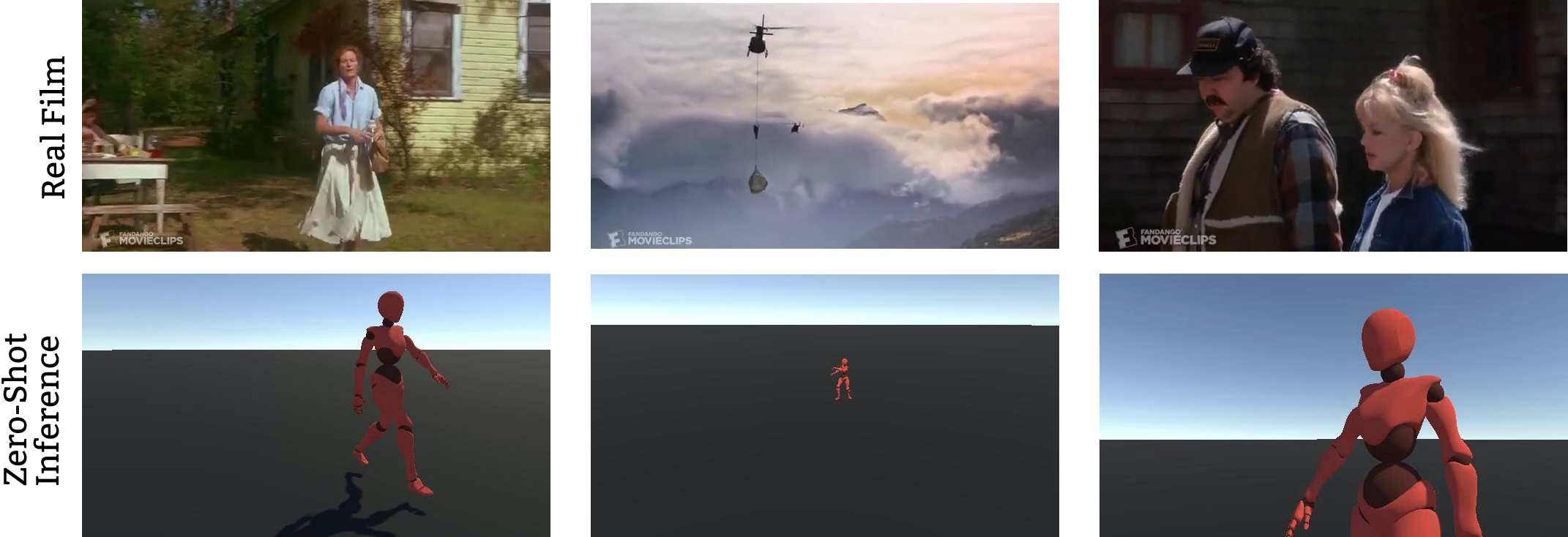}{\centering}
    \caption{Example-based camera motion generation with real films. The CLIP module also enables embedding real film images as conditions, allowing for the generation of camera sequences that exhibit cinematic framing similar to the provided reference frames.}
    \label{fig:real2clip}
\end{figure}

\begin{figure*}[ht]
    \centering
    \includegraphics[clip,width=\textwidth]{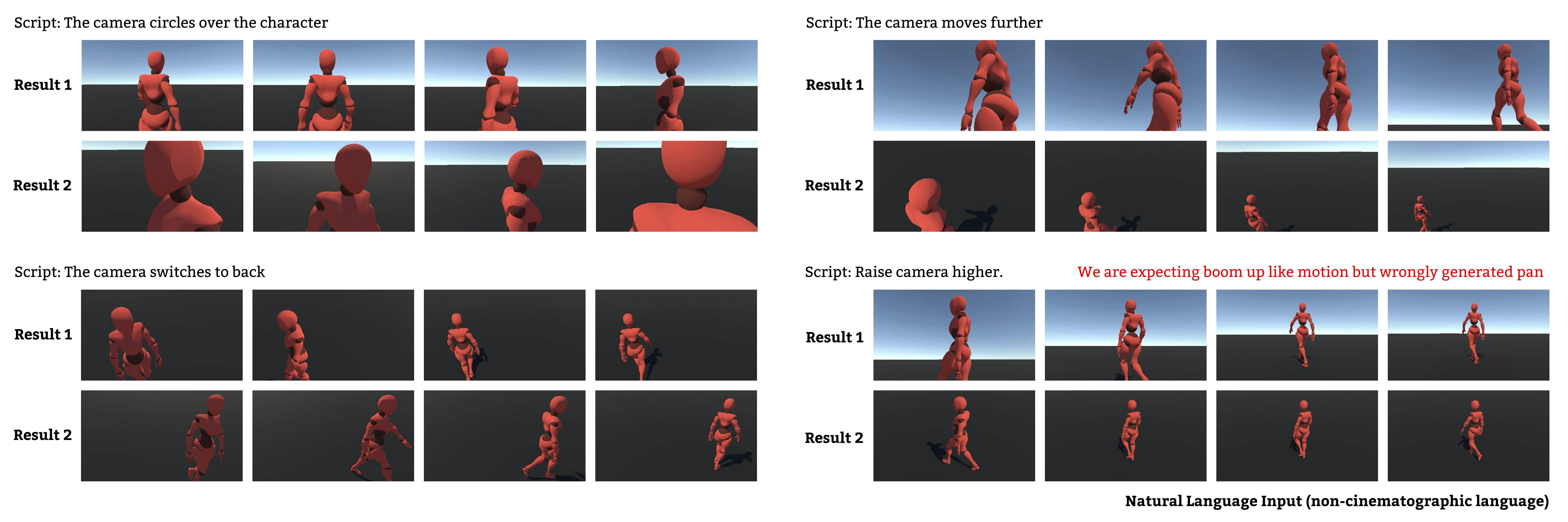}{\centering}
    \caption{Examples of casual language input, showing that our method can handle non-strict cinematographic language to a certain extent, though erroneous results can still occur in some cases.}
    \label{fig:zero-shot}
\end{figure*}

  
  

\begin{figure*}
  \centering

\includegraphics[clip,width=\textwidth]{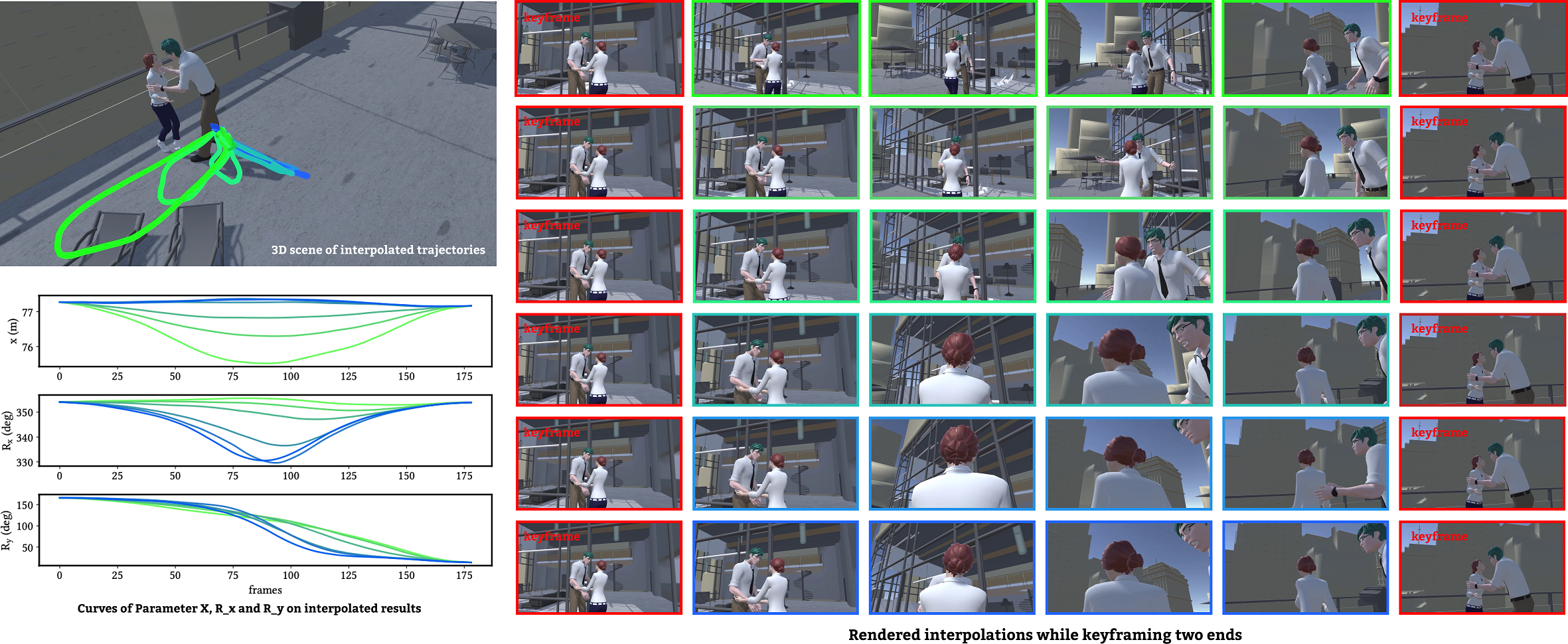}{\centering}
  
\caption{\textbf{Interpolation with keyframing.} Visualization of keyframe constrained CLIP interpolated results from \textit{orbit} (green) to \textit{pan} (blue) gradually, the interpolated results are achieved by linear interpolating (20\% per step) on the CLIP text embedding space while fixing two ends of keyframes.}
  \label{fig:transition_interpolation}
\end{figure*}

\begin{figure}[ht]
    \centering
    \includegraphics[clip,width=0.37\textwidth]{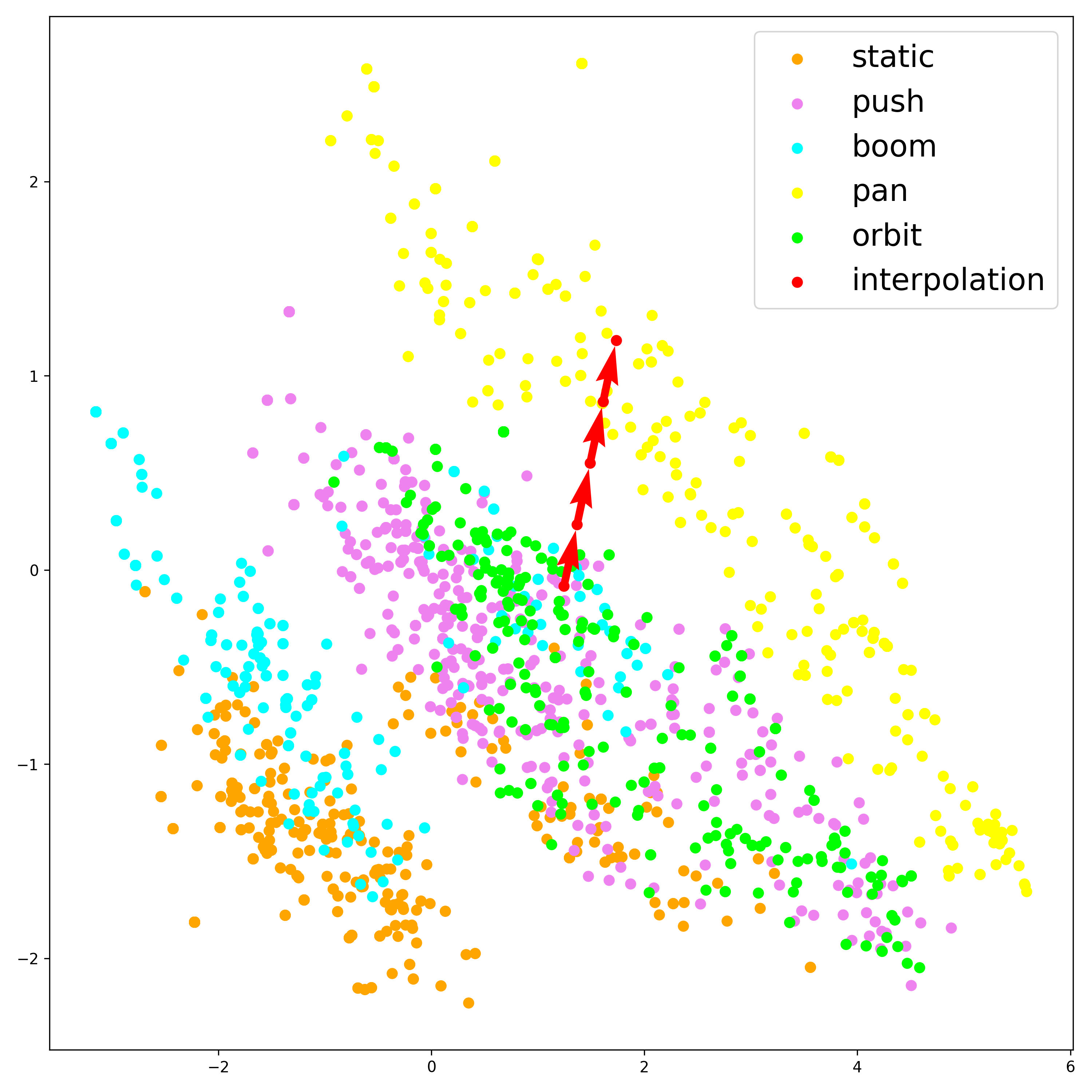}{\centering}
    \caption{Visualization of CLIP embeddings with different camera movements using PCA reduction. The interpolated embeddings between '\textit{pan}' and '\textit{orbit}' are represented by red quivers.}
    \label{fig:clip_viz}
\end{figure}

\begin{figure}[ht]
    \centering
    \includegraphics[width=0.48\textwidth]{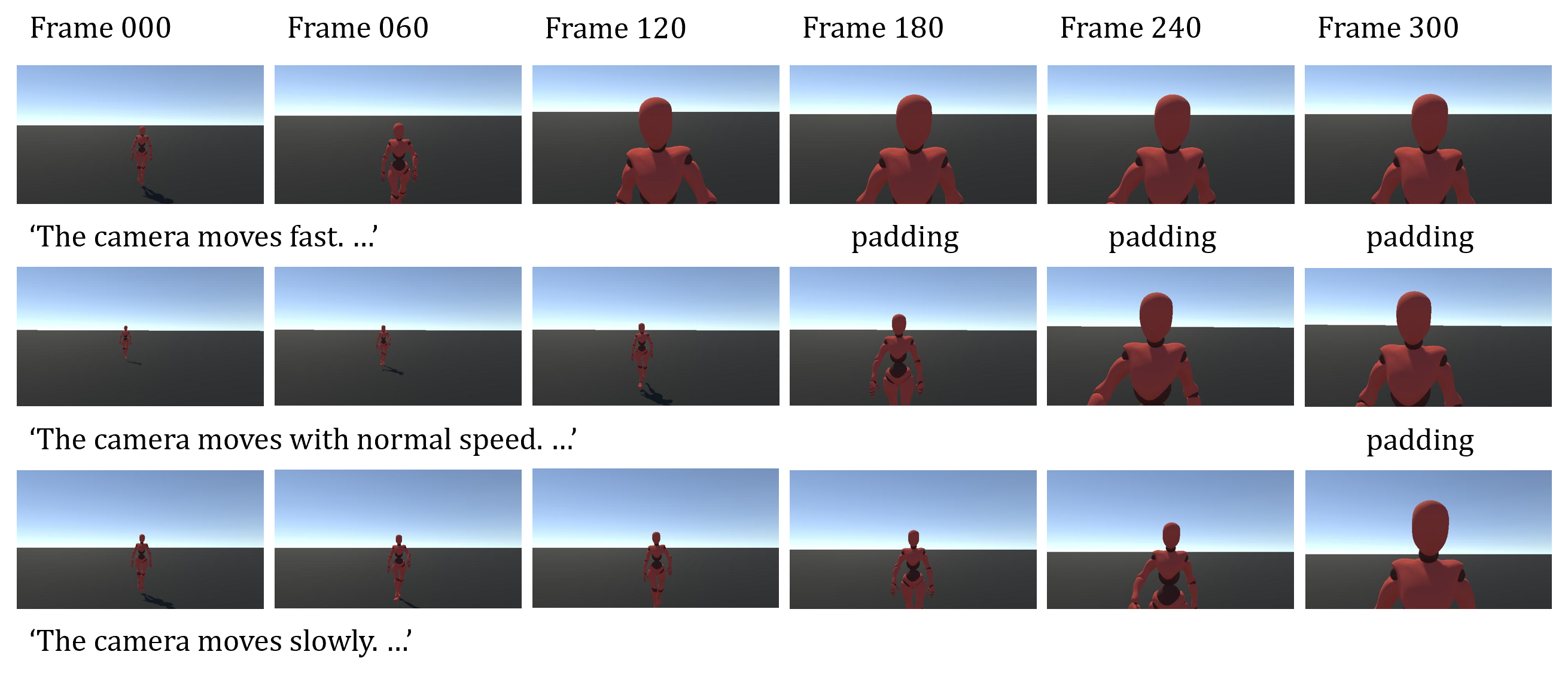}{\centering}
    \caption{Illustration of control on the camera velocity with different text prompts. Each sequence is generated with the same prompt '\textit{The camera pushes into the character. The camera shoots at eye level. The camera moves in front view. The character is at the middle center of the screen. The camera moves from long shot to close shot.}' Textual specifications related to different camera velocities are added at the beginning of these prompts, and the padded frames are indicated by the "padding".}
    \label{fig:vel_control}
\end{figure}

\begin{figure*}
  \centering
\includegraphics[clip,width=\textwidth]{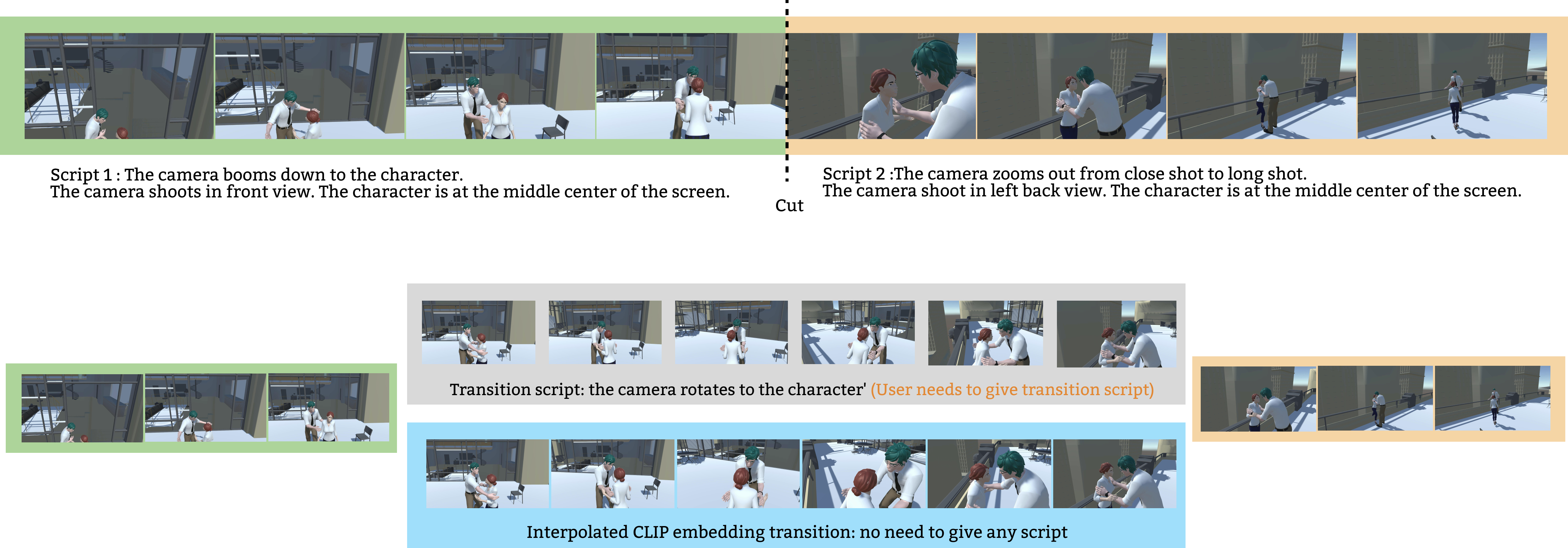}{\centering}
  
\caption{Comparison of transition schemes. Our proposed transition can smoothly connect different styles with no need for transition script. The interpolated CLIP embedding transition uses the average of the former and latter script embeddings as the text condition.}
  \label{fig:long_sequence}
\end{figure*}

\subsection{Qualitative studies and ablation}
\label{subsec:evaluation}


\subsubsection{Variation \emph{vs.} linguistic uncertainty of script}
\begin{figure*}[h!]
    \centering
    \includegraphics[clip,width=.9\textwidth]{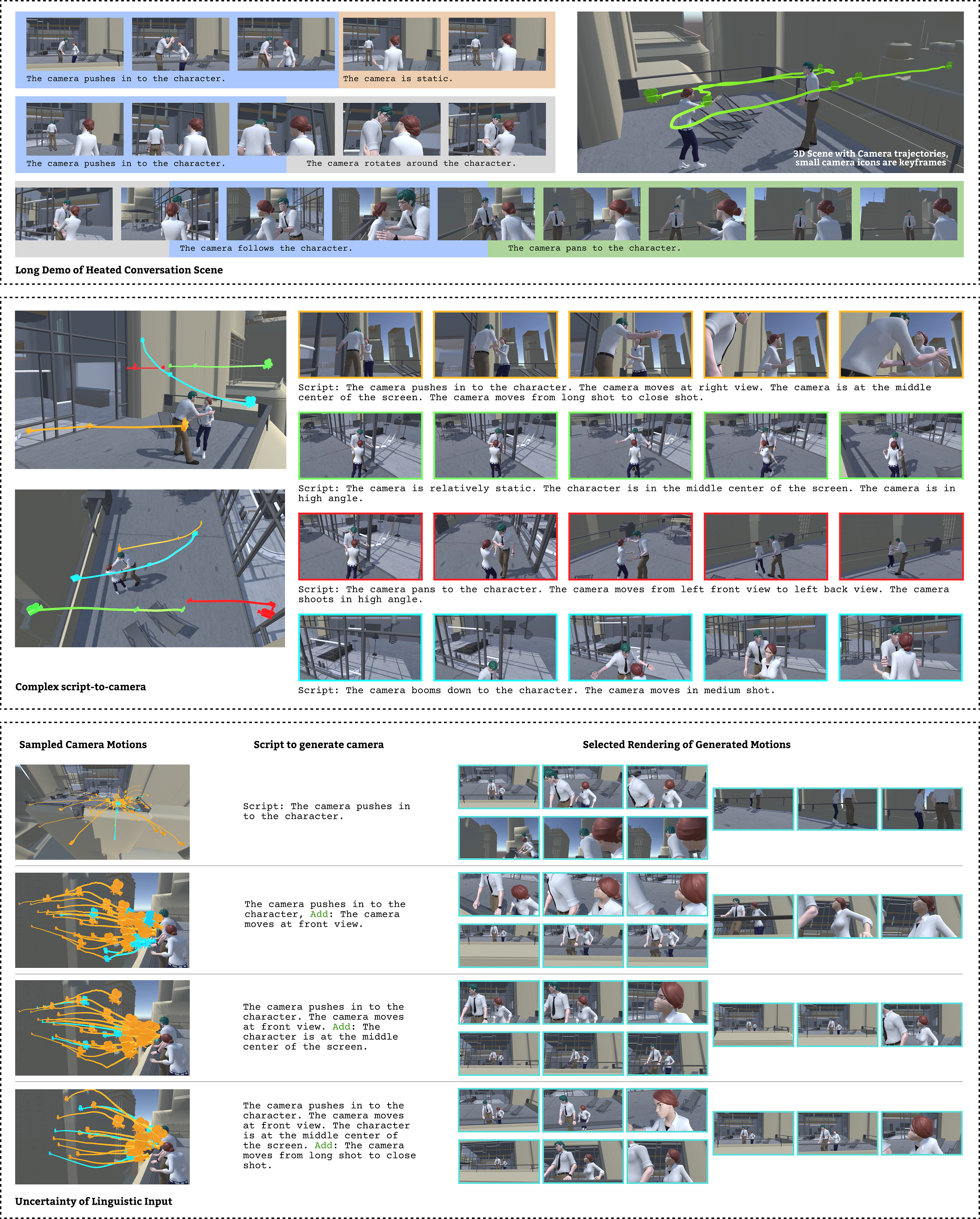}{\centering}
    \caption{Demonstrations of long video generation in various styles, showcasing of generation on complex script input, and visualizations of the impact of gradually adding linguistic descriptions during the generation process, highlighting the progression and refinement of the generated camera motions as more detailed prompts are provided.}
    \label{fig:all_image_page1}
\end{figure*}

To address the challenge of the many-to-many mapping from prompt to camera motion, we employ a prompt augmentation strategy during training to help generate diverse results for a given script prompt. In Fig.~\ref{fig:prompt_level}, we demonstrate the conflict between variation and linguistic uncertainty in script prompts, when provided with a vague script such as \textit{"The camera pushes in to the character"}, our model is capable of generating a multitude of results due to the stochasticity of the denoising process by sampling different noises. As more descriptive information is iteratively appended to the prompt, the uncertainty is reduced, leading to fine-grained and specific generated camera motions (see Fig.~\ref{fig:prompt_level}, Fig.~\ref{fig:all_image_page1} and supplementary video). This confirms two aspects: i) the existence of many-to-many ambiguities in the camera control task, where multiple camera motions can correspond to a single script prompt; and ii) with our model, users can achieve fine control and a wide range of variability by inputting different levels of precision in their scripts.

\subsubsection{Example-based generation}
CLIP embedding is renowned for its high discriminative capacity in handling linguistic features and for its ability to provide a joint embedding space between images and text, enabling zero-shot applications. We leverage this feature to demonstrate the multimodality potential of our proposed method. Inspired by~\cite{ao2023gesturediffuclip}, we utilize CLIP image embedding to process real film sequences frame by frame, followed by an average pooling layer. The resulting embeddings are then directly connected to our system, bypassing the text embedding input. Fig.~\ref{fig:real2clip} presents examples showcasing the capability of our method to capture certain cinematic information, such as screen position, framing, and camera motion, even under zero-shot multimodal conditions (see more in the supplementary video). However, it should be noted that due to the limited and noisy information (especially temporal) provided by individual frames, the camera motions generated from the CLIP image embeddings may not be as precise as those generated from script inputs.

\subsubsection{Zero-shot natural language generation}
One of the challenges in text-based generation tasks lies in the robustness to casual or non-strict text inputs. Through the use of prompt augmentation strategy, as well as leveraging the zero-shot capacity of CLIP features, we can handle various non-strict cinematographic language inputs never seen during the training \eg \textit{"camera switches to ...", "circling", "move ... further away"} and generate corresponding camera motions. Some examples are demonstrated in Fig.~\ref{fig:zero-shot}, yet it is important to note that erroneous cases can still occur, as CLIP embedding is not specifically influenced by "cinematic" language bias. Addressing this issue can be a direction for future research.

\subsubsection{Interpolation on CLIP space}
As mentioned in the previous section, our approach combines keyframing techniques with interpolation on CLIP text embeddings to achieve continuous and styled transition while retaining keyframing control. In Fig.~\ref{fig:transition_interpolation}, we provide an example of CLIP text embedding interpolation between the styles of \textit{'orbit'} and \textit{'pan'}, while keeping the keyframes fixed. In subfigure (a), we visualize the trajectories of the interpolated parameters $X$, $R_x$, and $R_y$. The color interpolation between blue (\textit{pan}) and green (\textit{orbit}) represents the different levels of linear interpolation on the CLIP text embeddings. We perform interpolation steps of 20\%, starting from $1 \times c_s(\text{\textit{orbit}})+0 \times c_s(\text{\textit{pan}})$ and gradually interpolating to $0.8 \times c_s(\text{\textit{orbit}})+0.2 \times c_s(\text{\textit{pan}})$, until reaching the all-\textit{pan} style (blue), with $c_s$ denotes the text CLIP embedding. We visualize the interpolated embeddings in Fig.~\ref{fig:clip_viz}. The results demonstrate that the camera trajectories smoothly transition between the two styles while maintaining the keyframing information.

\subsubsection{Transition schemes for long sequence generation}
Once the keyframing, script, and interpolated embedding control are in place, our method is capable of generating long sequences of cinematographic camera motions. We provide an example of a long sequence generated for a heated conversation scene in Fig.~\ref{fig:long_sequence}. By providing different camera scripts along with desired keyframing information and the duration of each motion, our method can generate long sequences of motions that are believable, styled, smoothly transitioned, and controllable.

In the supplementary video, we compare long sequences using different schemes. Given user-specified keyframes as constraints, only a few text prompts are needed to control the generation. When only generating sequences with text, long text prompts with different cinematic properties are required to generate a target sequence, and an automatic transition scheme is required to achieve continuous camera movement without jump cuts.

\subsubsection{Velocity control}
\label{label:vel_control}
To offer users the ability to control the camera velocity, we also added to the dataset a collection of motions with different camera velocities. Since we rely on sequences of fixed duration for the training, we simply padded the remaining frames with static camera poses for faster camera motions. Semantic annotations are based on measuring the speed of the camera along the trajectory. 
In Fig.~\ref{fig:vel_control}, we illustrate this velocity control during inference by generating various camera trajectories of different velocities, thereby influencing the temporal duration of the shots. With different text prompts, the camera moves from a 'long shot' to a 'close shot' within a shorter or longer duration. In the training phase, we use a padding strategy to achieve the target length for transformers. During inference, users control camera velocity through prompts and the model will generate the padding frames, which can subsequently be truncated.



\subsubsection{Influence of $\omega$}
\label{subsection:omega}
We show the ablation of the guidance scale with near-style keyframing experiments in Tab.~\ref{tab:guidance_scale}. We evaluate our model with $\omega=1$ as the accuracy-fidelity sweet spot.

\begin{table}[h!]
    \centering
    \caption{We evaluate how different classifier-free guidance scale $\omega$ affect the performance of proposed method.}
    \begin{tabular}{l|c|c|c}
        \hline
            & \multicolumn{3}{|c}{Near-style KF} \\
        \hline
            Method & FID$\downarrow$ & KD $\downarrow$ & R Prec $\uparrow$ \\
        \hline
        Ours ($\omega$=0.5) & 7.01 & \textbf{0.040} & 95.71  \\
         Ours ($\omega$=1) & \textbf{6.67} & 0.041 & \textbf{96.62} \\
         Ours ($\omega$=1.5) & 7.63 & 0.045 & 94.99 \\
        \hline
    \end{tabular}
    \label{tab:guidance_scale}
\end{table}

\subsubsection{Experts evaluation}

After sharing our results with artists (x2), film directors and directors of photography (x2), and producers (x1) in the film and animation industries, we have summarized their reactions (the raw transcripts are listed in the supplementary document).

\noindent
\textbf{Potential}: Almost every participant expressed their belief in the high potential of the proposed method for the future combination of AI and filmmaking. Some mentioned the integration of metahumans and Large Language Models as potential avenues for exploration.

\noindent
\textbf{Non-professional and professional application}: Many participants noted that the tool could be used for non-professional purposes, such as experiencing the filmmaking process in platforms like TikTok or virtual reality. For professionals, the tool could be utilized to replace some human labor, particularly in previz, facilitating intuitive communication with directors to staff, and capturing non-complex routine shots.

\noindent
\textbf{Efficiency and automation}: The tool's pros were seen in its efficiency and automation, allowing it to reach a comparable level to a mediocre-pro cameraman, and find shooting problems earlier. Issues were perceived with sometimes too fast orbit motions with rotating characters. Overall the efficiency was appreciated by participants.

\noindent
\textbf{Creative limitations and systematic decision-making}: Some participants raised concerns about the creative limitations of the learned patterns, suggesting that the tool may restrict users' creative freedom. Others mentioned that addressing camera motion as a standalone tool might not be sufficient, as the entire filmmaking process requires systemic and integrated decision-making, encompassing shooting, environment, and mise-en-scene simultaneously.

%% file: 7_conclusion.tex
\subsection{Limitation}
\textbf{Semantic language input}: Current methodologies lack in incorporating rich semantic language inputs that include context and emotional correspondence aligned with real film shooting data. Enhancing the semantic depth of language inputs could significantly improve the realism and applicability of generated scenes.

\noindent
\textbf{Character-centric coordinate system}: While the character-centric coordinate system effectively addresses the character-camera pairing challenge, ensuring the camera's focus on the character, it introduces potential drawbacks. These include: i) it is prone to generating shaky camera motions, especially when the character motion is active; ii) a limitation in exploring more complex compositional possibilities. E.g., it hinders the depiction of complicated movements involving multiple characters, such as long shots with complex trajectories; iii) difficulties in capturing scenes where the character is absent, which are common in various cinematic contexts like scene touring, commercials, and shots that establish context, restricting the narrative and visual storytelling capabilities.

\subsection{Conclusion}
In this paper, we presented a diffusion approach to generate realistic camera motions relative to targets. We showed how to address the specific requirements in camera motion generation techniques, such as transitioning between motions using embedding interpolation, accounting for keyframe conditions to improve user control, or augmenting the prompt diversity. The stochasticity of the diffusion techniques helps to generate convincing results with partial textual descriptions and displays the benefits of the
approach over recent techniques. The work presents some limitations such as accounting for precise timings, dealing with sequential movements, and lacking elaborate modifiers on the textual specifications (faster/smoother/...) mostly due to the lack of available data. The ongoing challenges consist of exploring how such textual descriptions can be obtained automatically from real film clips with more precision, accounting for a far richer dataset and therefore a wider variability in generated results with more elaborate modifiers.

%% file: 9_ack.tex
We want to thank Anthony Mirabile for his valuable support throughout this project. We also wish to thank the anonymous reviewers for their constructive comments. This work was in part supported by the National Key R\&D Program of China 2022ZD0160803.

%% file: 10_appendix.tex
\begin{figure}[ht]
    \centering
    \includegraphics[width=0.3\textwidth]{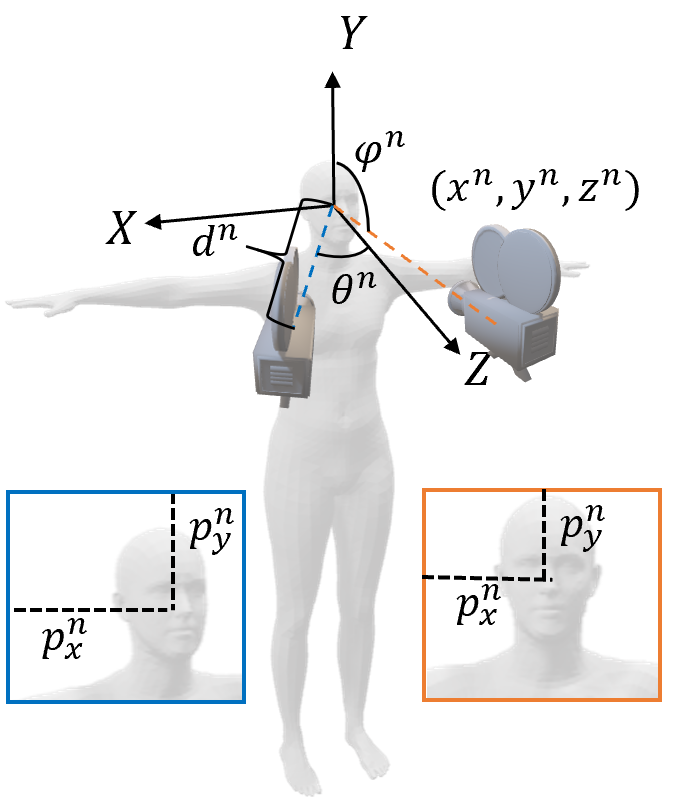}{\centering}
    \caption{Illustration of the character-centric local coordinate system. The system is built under a left-handed coordinate system, where the $\{X, Y, Z\}$ axes represent the left, up, and forward directions, respectively.}
    \label{fig:appendix_cam}
\end{figure}

\section{Dataset}
\label{appendix:data_gen}

\subsection{Camera representation and prompt templates}



Our camera trajectories are formulated as a list of 5DoF camera parameters: $\mathbf{c}=\{(x^n,y^n,z^n),(p_x^n,p_y^n)\}$ (see Section~\ref{label:Method}). For convenience, we also use additional variables $\{\theta^n, \varphi^n, d^n\}$ to express the camera in a local spherical coordinate system (elevation, azimuth, and distance $d^n$), as shown in Fig.~\ref{fig:appendix_cam}. Here we provide detailed information on how these parameters are related to different cinematic properties, along with example prompt templates.

\begin{itemize}
\item Shot angle
    \begin{itemize}
        \item $\varphi^n \in \{[0.25\pi,0.4\pi], (0.4\pi,0.6\pi), [0.6\pi,0.75\pi]\}$ represents \textit{high-angle, eye-level and low angle}.
        \item \textit{The camera shoots the character at <shot angle>.}
        \item \textit{The camera shoots in <shot angle>.}
    \end{itemize}
\item Shot scale
    \begin{itemize}
        \item $d^n \in \{[0.1h,0.2h),[0.2h,0.3h),[0.3h,0.6h),[0.6h,1.2h)$, $[1.2h,2.0h),[2.0h,4.0h)\}$ represents \textit{extreme close, close, medium close-up, medium, long, extreme long shots}, where $h$ is the height of the character.
        \item \textit{The camera shoots at <shot scale>.}
        \item \textit{The camera moves from <starting shot scale> to <ending shot scale>.}
    \end{itemize}
\item View directions
    \begin{itemize}
        \item $\theta^n\in \{0,0.25\pi,0.5\pi,0.75\pi,\pi,1.25\pi,1.5\pi,1.75\pi\}$ represents \textit{front, right front, right, right back, back, left back, left, left front}. For each direction, $\theta^n$ has a  variation with $\pm 0.05\pi$.
        \item \textit{The camera shoots in <view direction>.}
        \item \textit{The camera switches from <starting view direction> to <ending view direction>.}
    \end{itemize}
\item Screen properties
    \begin{itemize}
        \item The screen coordinate system is normalized to $[-1,1]\times[-1,1]$. $p_x^n \in \{[-0.7,-0.3],[-0.2,0.2],[0.3,0.7]\}$ represents horizontal position \textit{left, center, right} and $p_y^n \in \{[-0.7,-0.3],[-0.2,0.2],[0.3,0.7]\}$ represents vertical position \textit{top, middle, bottom}.
        \item \textit{The character is at the <screen property> of the screen.}
    \end{itemize}
\item Shot movement
    \begin{itemize}
        \item $\{(x^n,y^n,z^n),(p_x^n,p_y^n)\}=\{(x^0,y^0,z^0),(p_x^0,p_y^0)\}$ represents \textit{static}, where the camera remains locally static in both position and framing. 
        \item $\{(x^n,y^n,z^n),(p_x^n,p_y^n)\}=\{(x^0,y^0,z^0) * d^n / d^0,(p_x^0,p_y^0)\}$ represents \textit{push in/pull out}, where the camera either decreases or increases the distance $d^n$ to the character while keeping other properties constant.
        \item $\{(x^n,y^n,z^n),(p_x^n,p_y^n)\}=\{(x^0,y^0,z^0+n*\mathrm{d}z),(p_x^0,p_y^0)\}$ represents \textit{pan}, where the screen properties remains constant while it rotates on its vertical axis to track a target.
        \item $\{(x^n,y^n,z^n),(p_x^n,p_y^n)\}=\{(x^0,y^0+n*\mathrm{d}y,z^0),(p_x^0,p_y^0+n*\mathrm{d}p_y)\}$ represents \textit{boom}, where $\frac{\mathrm{d}y}{\mathrm{d}p_y} = d^0$. A boom motion typically involves an upward or downward translational movement of the camera.
        \item $\{(x^n,y^n,z^n),(p_x^n,p_y^n)\}=\{(\sin\theta^n d^0,y^0,\cos\theta^n d^0),(p_x^0,p_y^0)\}$ where $\theta^n=\theta^0+n*\mathrm{d}\theta$ represents \textit{Orbit}, where the camera swivels horizontally around a target with angular velocity $\mathrm{d}\theta$.
        \item \textit{The camera is static.}
        \item \textit{The camera pushes in/out to the character.}
        \item \textit{The camera pans to the character.}
        \item \textit{The camera booms up/down.}
        \item \textit{The camera rotates around the character.}
    \end{itemize}
\item Shot velocity
    \begin{itemize}
        \item $n \in \{[75,105], [160,200], [240,300]\}$ with fps 30 represents 'fast, normal, and slow'.
        \item \textit{The camera moves <velocity>.}
    \end{itemize}
\end{itemize}

\section{Network structure}
\label{appendix:network}

The full architecture of the classification network is summarized in the table below, where \texttt{FC}, \texttt{TE}, \texttt{ReLU}, and \texttt{L} denote the fully connected layer, transformer encoder layer, ReLU activation layer, and the length of data, respectively. The number of input and output channels are reported in the rightmost column. The network is optimized with a cross-entropy loss implemented with PyTorch.

\begin{center}
\begingroup
    \fontsize{8pt}{10pt}\selectfont

  \begin{tabular}{| l | l | l | c |}
  \hline
      Name & Modules & Layers  & in/out \\
  \hline
  \hline
      Camera & Input process & \texttt{FC}  & \texttt{L}*5/\texttt{L}*256\\
      Classification & Positional encoding & - & \texttt{L}*256/\texttt{L}*256\\
      Network & Transformer & \texttt{TE}*6 & \texttt{L}*256/\texttt{L}*256\\
      & Predictor & \texttt{FC+ReLU+FC} & \texttt{L}*256/\texttt{L}*6\\
  \hline
\end{tabular}
\end{center}
\endgroup